%% file: circle_graph_iso.tex
\documentclass[11pt,a4paper]{article}
\usepackage{authblk}
\usepackage{amssymb,amsmath,amsfonts,amsthm}
\usepackage[a4paper,margin=2.5cm]{geometry}
\usepackage{hyperref}
\usepackage[colorinlistoftodos,backgroundcolor=blue!10,linecolor=red,bordercolor=red]{todonotes}
\usepackage{xspace}
\usepackage{graphicx}

\newtheorem{theorem}{Theorem}[section]
\newtheorem{lemma}[theorem]{Lemma}

\theoremstyle{definition}
\newtheorem{definition}[theorem]{Definition}

\newtheorem{problem}[theorem]{Problem}

\theoremstyle{remark}

\numberwithin{equation}{section}

  \def\calC{{\mathcal C}}
 \def\calE{{\mathcal E}}

  \def\calO{{\mathcal O}}
  \def\calR{{\mathcal R}}
 \def\calT{{\mathcal T}}

\def\cNP{\hbox{\rm \sffamily NP}\xspace}

\def\eps{\varepsilon}
\def\O{\mathcal{O}{}}

\def\inter#1{\left<#1\right>}

\def\O{\mathcal{O}{}}

\def\int{\hbox{\rm \sffamily INT}\xspace}

\def\path{\hbox{\rm \sffamily PATH}\xspace}

\def\eps{\varepsilon}
\def\dotcup{\mathbin{\dot\cup}}

\def\O{\calO}

\title{Circle Graph Isomorphism in Almost Linear Time}

\author[1]{V\'{\i}t Kalisz}
\author[2]{Pavel Klav\'{\i}k}
\author[1]{Peter Zeman}

\affil[1]{Department of Applied Mathematics,\protect\\
Faculty of Mathematics and Physics,\protect\\
Charles University, Prague, Czech Republic,\protect\\
\texttt{vitek.kalisz@gmail.com}, \texttt{zeman@kam.mff.cuni.cz}}
\affil[2]{Computer Science Institute,\protect\\
Faculty of Mathematics and Physics,\protect\\
Charles University, Prague, Czech Republic,\protect\\
\texttt{klavik@iuuk.mff.cuni.cz}}

\date{}

\begin{document}

\maketitle

\begin{abstract}
Circle graphs are intersection graphs of chords of a circle.  In this paper, we present a new
algorithm for the circle graph isomorphism problem running in time $\O((n+m)\alpha(n+m))$ where $n$
is the number of vertices, $m$ is the number of edges and $\alpha$ is the inverse Ackermann
function. Our algorithm is based on the minimal split decomposition [Cunnigham, 1982] and uses the
state-of-art circle graph recognition algorithm [Gioan, Paul, Tedder, Corneil, 2014] in the same
running time. It improves the running time $\O(nm)$ of the previous algorithm [Hsu, 1995] based on
a similar approach.
\end{abstract}


\input introduction
\input split_decomposition

\input canonization_of_trees
\input canonization_of_nodes

\input canonization_of_circle_graphs
\input conclusions

\bibliographystyle{plain}
\bibliography{circle_graph_iso}

\end{document}

%% file: introduction.tex
\section{Introduction} \label{sec:introduction}

For graphs $G$ and $H$, a bijection $\pi : V(G) \to V(H)$ is called an \emph{isomorphism} if $uv \in
E(G) \iff \pi(u)\pi(v) \in E(H)$. Testing isomorphism of graphs in polynomial time is a major open
problem in theoretical computer science. The graph isomorphism problem clearly belongs to \cNP, it is
unlikely \cNP-complete~\cite{schoning1988graph}, and the best known algorithm runs in quasipolynomial
time~\cite{babai_quasipoly}, heavily based on group theory. For a survey, see~\cite{kkz}.

For restricted classes of graphs, it is often possible to design combinatorial polynomial-time
algorithms, relying on strong structural properties of the considered classes: prime examples are
linear-time algorithms for testing isomorphism of trees~\cite{aho_hopcroft_ullman} and planar
graphs~\cite{hopcroft1974linear,fkkn16,kkmnz}. Testing isomorphism of interval
graphs~\cite{lueker1979linear} and permutation graphs~\cite{permutation_isomorphism,kz} in linear
time relies on the property that certain canonical trees describe all their representations, and
thus it can be reduced to testing isomorphism of trees whose nodes are labeled by simple graphs. In
this paper, we described an almost linear-time algorithm for testing isomorphism of circle graphs
working in a similar spirit.

\begin{figure}[b]
\centering
\includegraphics{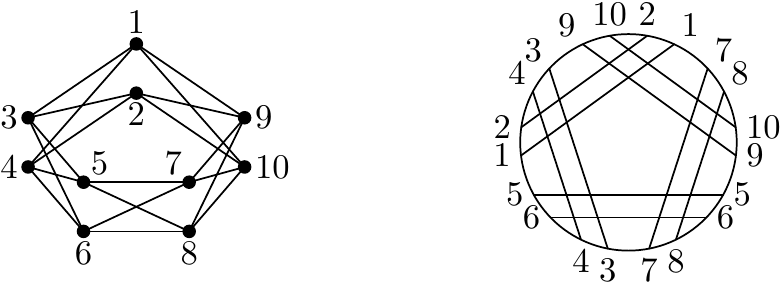}
\caption{A circle graph and one of its circle representations corresponding to the circular word
$10,2,1,7,8,10,9,5,6,8,7,3,4,6,5,1,2,4,3,9$.}
\label{fig:introduction_example}
\end{figure}

For a graph $G$, a \emph{circle representation} $\calR$ of a graph $G$ is a collection of sets
$\bigl\{\inter v : v \in V(G)\bigr\}$ such that each $\inter v$ is a chord of some fixed circle, and
$\inter u \cap \inter v \neq \emptyset$ if and only if $uv \in E(G)$. Observe that $\calR$ is
determined by the circular word giving the clockwise order of endpoints of the chords in which $uv
\in E(G)$ if and only if their endpoints alternate as $uvuv$ in this word. A graph is called a
\emph{circle graph} if and only if it has a circle representation; see
Fig.~\ref{fig:introduction_example}.

Circle graphs were introduced by Even and Itai~\cite{EI71} in the early 1970s.  They are related to
Gauss words~\cite{deFM99}, matroid representations~\cite{B87,deF81}, and rank-width~\cite{oum}.  The
complexity of recognition of circle graphs was a long-standing open problem, resolved in
mid-1980s~\cite{B1987,GSH89,N85}. Currently, the fastest recognition algorithm~\cite{GPTC13} runs in
almost linear time. In this paper, we use this recognition algorithm as a subroutine and solve the graph
isomorphism problem of circle graphs in the same running time.

\begin{theorem} \label{thm:graphiso}
The graph isomorphism problem of circle graphs and the canonization problem of circle graphs can be
solved in time $\calO((n+m) \cdot \alpha(n+m))$, where $n$ is the number of vertices, $m$ is the
number of edges, and $\alpha$ is the inverse Ackermann function. Further, if circle
representations are given as a part of the input, the running time improves to $\calO(n+m)$.
\end{theorem}

Two circle representation are isomorphic if by relabeling the endpoints we get identical circular
orderings. In Section~\ref{sec:canonization_of_nodes}, we show that isomorphism of circle
representations can be tested in time $\calO(n)$. When circle graphs $G$ and $H$ have isomorphic circle
representations $\calR_G$ and $\calR_H$, clearly $G \cong H$. But in general, the converse does not
hold since a circle graph may have many non-isomorphic circle representations.

The main tool is the split decomposition which is a recursive process decomposing a graph into
several indecomposable graphs called \emph{prime graphs}. Each split decomposition can be described
by a split tree whose nodes are the prime graphs on which the decomposition terminates. The key
property is that the initial graph is a circle graph if and only if all prime graphs are circle
graphs. Further, each prime circle graph has a unique representation up to reversal, so
isomorphism for them can be tested in $\calO(n)$ using Section~\ref{sec:canonization_of_nodes}.

One might want to reduce the isomorphism problem of circle graphs to the isomorphism problem of
split trees. Unfortunately, a graph may posses many different split decompositions corresponding to
non-isomorphic split trees. The seminal paper of Cunnigham~\cite[Theorem
3]{cunningham1982decomposition} shows that for every connected graph, there exists a minimal split
decomposition; this result was also proven in~\cite[Theorem 11]{cunningham1980combinatorial}.
The split tree associated to the minimal split decomposition is then also unique and
it follows that the isomorphism problem of circle graphs reduces to the isomorphism problem of minimal split trees.

Already in 1995, this approach was used by Hsu~\cite{hsu1995m} to solve the graph isomorphism
problem of circle graphs in time $\calO(nm)$. He actually concentrates on circular-arc graphs, which
are intersection graphs of circular arcs, and builds a decomposition technique which generalizes
minimal split decomposition. The main results are recognition and graph isomorphism algorithms for
circular-arc graphs running in $\calO(nm)$. Unfortunately, a mistake in this general decomposition
technique was pointed out by~\cite{isomorphism_circular_one_property}.

This discovered mistake does not affect the graph isomorphism algorithm for circle graphs. However,
Hsu's 29 page long paper~\cite{hsu1995m} just briefly mentions circle graphs and minimal split
decomposition, with many details omitted.
Also, a straightforward adaptation of Hsu's algorithm combined with the fastest
algorithm computing the minimal split decomposition implies an $\calO(n^2)$ algorithm for testing
isomorphism of circle graphs. But to get almost linear running time, further problems have to be
addressed in this paper.

In~\cite{lin2008simple}, a linear-time algorithm testing isomorphism of \emph{proper circular-arc
graphs} is described. These are intersection graphs of circular arcs such that no arc is properly
contained in another one. Since proper circular-arc graphs form a subclass of circle graphs, we
strengthen their result.

\subparagraph*{Outline.}
Section~\ref{sec:split_decomposition} gives an overview of minimal split decomposition and minimal
split trees. In Section~\ref{sec:canonization_of_trees}, we describe a meta-algorithm computing
cannonical form of a general tree whose nodes are labeled by graphs for which a linear-time
cannonization is known.  For a circle graph, its unique minimal split tree is labeled by prime
circle graphs and degenerate graphs (complete graphs and stars). In
Section~\ref{sec:canonization_of_nodes}, we give linear-time cannonization algorithms for prime and
degenerate circle graphs. By putting this together in
Section~\ref{sec:canonization_of_circle_graphs}, we prove Theorem~\ref{thm:graphiso}. In
Conclusions, we discuss related results and open problems.

%% file: split_decomposition.tex
\section{Minimal Split Decomposition and Split Trees} \label{sec:split_decomposition}

In this section, we describe several known properties of split decompositions and split trees.  We assume
that all graphs are connected, otherwise split decomposition is applied independently on each
component.

\begin{figure}[t]
\centering
\includegraphics{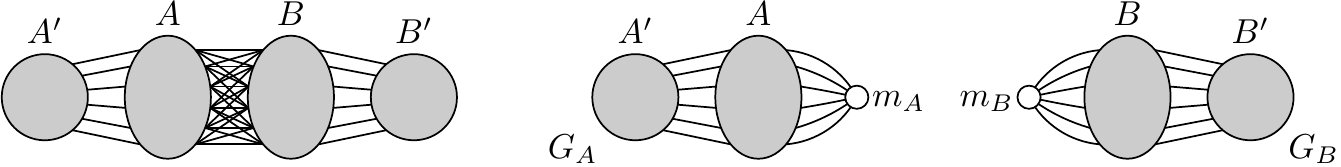}
\caption{On the left, a split in $G$ between $A$ and $B$. On the right, application of this split
produces graphs $G_A$ and $G_B$ with newly created marker vertices denoted by big white circles.}
\label{fig:split}
\end{figure}

\subparagraph*{Splits.}
For a graph $G$, a \emph{split} is a partition $(A,B,A',B')$ of $V(G)$ such that:
\begin{itemize}
\item For every $a \in A$ and $b \in B$, we have $ab \in E(G)$.
\item There are no edges between $A'$ and $B \cup B'$, and between $B'$ and $A \cup A'$.
\item Both sides have at least two vertices: $|A \cup A'| \ge 2$ and $|B \cup B'| \ge 2$.
\end{itemize}
See Fig.~\ref{fig:split} on the left. In other words, the cut between $A$ and $B$
is the complete bipartite graph. A split in a graph can be found in polynomial time~\cite{S94}.
Graphs containing no splits are called \emph{prime graphs}. Since the sets $A$ and $B$ already
uniquely determine the split, we call it the \emph{split between $A$ and $B$}.

We can \emph{apply} a split between $A$ and $B$ to divide the graph $G$ into two graphs $G_A$ and $G_B$
defined as follows. The graph $G_A$ is created from $G[A \cup A']$ together with a new \emph{marker
vertex} $m_A$ adjacent exactly to the vertices in $A$. The graph $G_B$ is defined analogously for
$B$, $B'$ and $m_B$. See Fig.~\ref{fig:split} on the right.

\subparagraph*{Split Decomposition and Split Trees.}
A split decomposition $D$ of $G$ is a sequence of splits defined as follows. At the beginning, we
start with the graph $G$. In the $k$-th step, we have graphs $G_1,\dots,G_k$ and we apply a split
on some $G_i$, dividing it into two graphs $G'_i$ and $G''_i$. The next step then applies to one of
the graphs $G_1,\dots,G_{i-1},G'_i,G''_i,G_{i+1},\dots,G_k$, and so on.

A split decomposition can be captured by a graph-labeled tree $T$. The vertices of $T$ are called
nodes to distinguish them from the vertices of $G$ and from the added marker vertices, and nodes
correspond to subsets of these vertices. To simplify the definition of graph isomorphism of
graph-labeled trees, we give a slightly different formal definition in which $T$ is not necessarily
a tree.

\begin{definition}
A graph-labeled tree $T$ is a graph $(V,E)$ with $E=E_N \dotcup E_T$ where $E_N$ are called
the \emph{normal edges} and $E_T$ are called the \emph{tree edges}. A \emph{node} is a connected
component of $(V,E_N)$. There are no tree edges between the vertices of one node and no vertex is
incident to two tree edges. The incidance graph of nodes must form a tree.
The \emph{size} of $T$ is $|V| + |E|$.
\end{definition}

\noindent A graph-labeled tree $T$ might not be a tree, but the underlying structure of tree edges
forms a tree of nodes. The vertices of $T$ incident to tree edges are called \emph{marker vertices}.

\begin{figure}[p!]
\centering
\includegraphics[width=\textwidth]{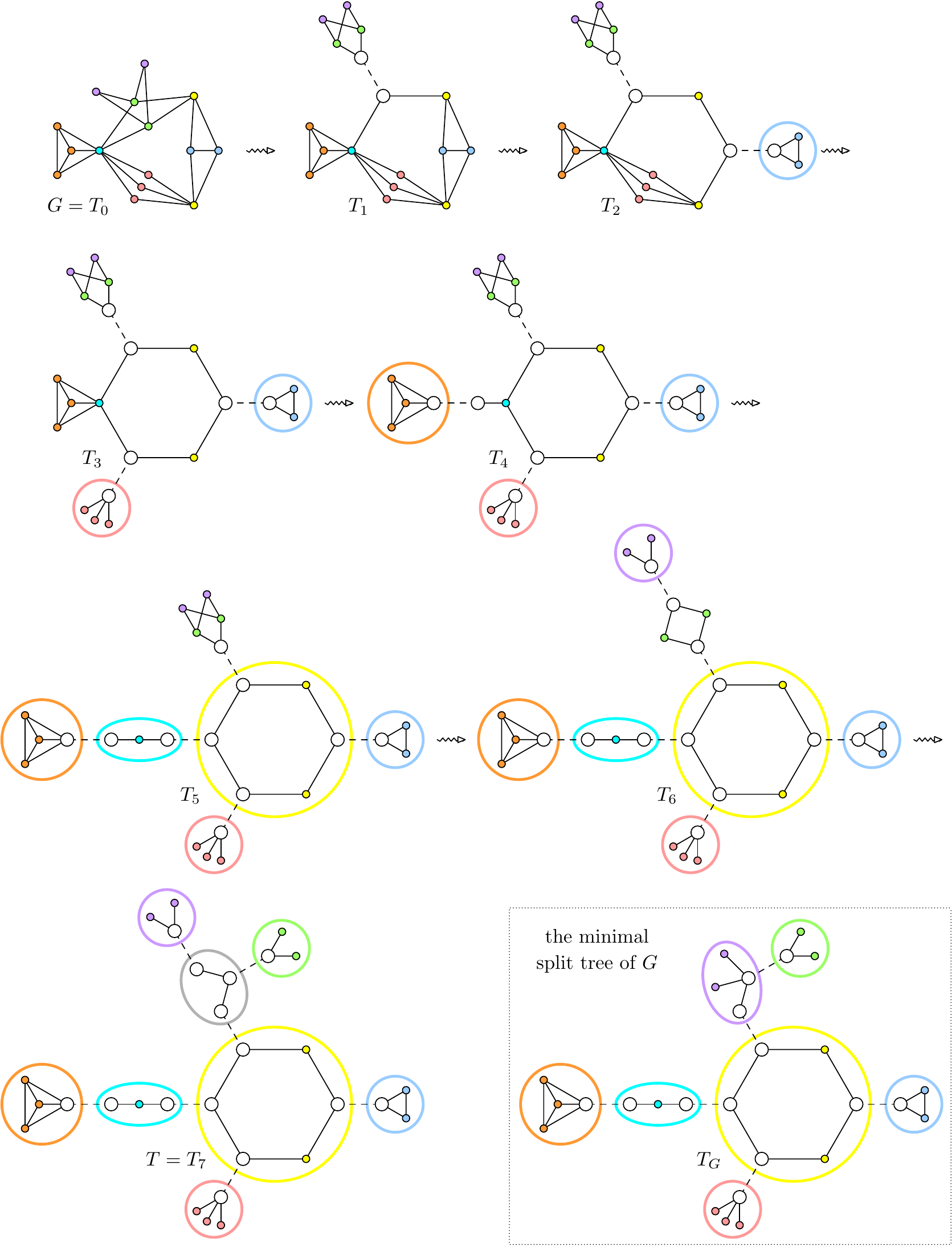}
\caption{An example of a split tree $T$ of a split decomposition $D$ ending with highlighted prime and
degenerate graphs (see the definition below). The split decomposition $D$ is not minimal: the gray
and purple stars can be joined in $T$ to form the minimal split tree $T_G$ in the box.}
\label{fig:split_tree}
\end{figure}

A split decomposition $D$ of $G$ is represented by the following graph-labeled tree $T$ called the
\emph{split tree $T$ of $D$} (or a split tree $T$ of $G$). Initially, $T$ consists of a single node
equal to $G$.  At each step, $D$ applies a split on one node $N$ of $T$. This node is replaced by
two new nodes $N_A$ and $N_B$ while the tree edges incident to $N$ are preserved in $N_A$ and $N_B$
and the marker vertices $m_A$ and $m_B$ are further adjacent by a newly formed tree edge.
Figure~\ref{fig:split_tree} shows an example. It can be observed that the total size of every split
tree is $\calO(n+m)$ where $n$ is the number of vertices and $m$ is the number of edges of the original
graph.

From a split tree $T$, the original graph $G$ can be reconstructed by \emph{joining neighboring
nodes}.  For a tree edge $m_Am_B$, we remove $m_A$ and $m_B$ while adding all edges $uv$ for $u \in
N(m_A)$ and $v \in N(m_B)$.

\subparagraph*{Recognition of Circle Graphs.}
A split decomposition can be applied to recognize circle graphs.
The key is the following observation.

\begin{lemma}
A graph is a circle graph if and only if both $G_A$ and $G_B$ are circle graphs.
\end{lemma}

\begin{figure}[t]
\centering
\includegraphics{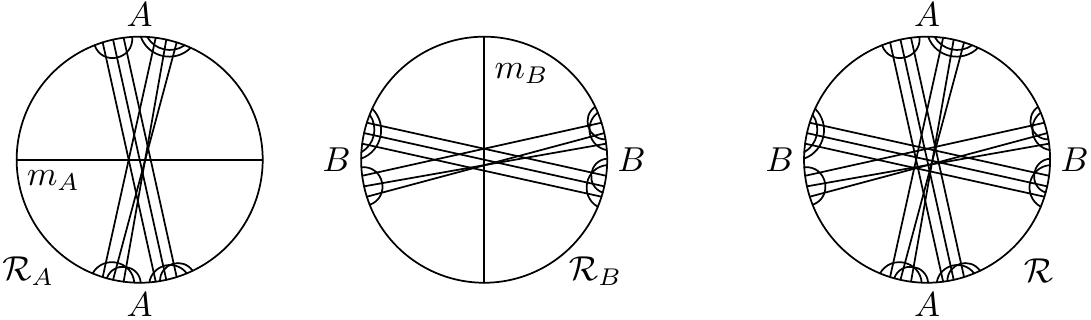}
\caption{On the left, circle representations $\calR_A$ and $\calR_B$ of graphs $G_A$ and $G_B$. They
are combined into a circle representation $\calR$ of $G$.}
\label{fig:recursion_on_split}
\end{figure}

The proof is illustrated in Figure~\ref{fig:recursion_on_split}, which can be easily formalized; see for example~\cite{egr}.
A prime circle graph has a unique circle representation up to reversal~\cite{dahlhaus} which can be constructed in polynomial time~\cite{GPTC13}.

\subparagraph*{Minimal Split Decomposition.} A graph is called \emph{degenerate} if it is the complete
graph $K_n$ or the star $S_n$. Suppose that we have a split decomposition $D$ ending on prime
graphs. Its split tree is not uniquely determined, for instance degenerate graphs
have many different split trees.  Cunnigham~\cite{cunningham1982decomposition} resolved this
issue by terminating the split decomposition not only on prime graphs, but also on the degenerate graphs.

Cunnigham~\cite{cunningham1982decomposition} introduced the notion of a \emph{minimal split
decomposition}. A split decomposition is minimal if the corresponding split tree has all nodes as
prime and degenerate graphs, and joining any two neighboring nodes creates a non-degenerate graph.

\begin{theorem}[Cunningham~\cite{cunningham1982decomposition}, Theorem 3]
\label{thm:minimal_decomposition}
For a connected graph $G$, the split tree of a minimal split decomposition terminating on prime and
degenerate graphs is uniquely determined.
\end{theorem}

\noindent The split tree of a minimal split decomposition of $G$ is called the \emph{minimal split
tree} of $G$ and it is denoted $T_G$.

It was stated in~\cite[Theorem 2.17]{GPTC13b} that a split decomposition is minimal if the
corresponding split tree has no two neighboring nodes such that
\begin{itemize}
\item either both are complete,
\item or both are stars and the tree edge joining these stars is incident to exactly one of their
central vertices.
\end{itemize}
The reason is that such neighboring nodes can be joined into a complete node or a star node,
respectively. Therefore, the minimal split tree can be constructed from an arbitrary split
tree by joining neighboring complete graphs and stars. For instance, the split tree $T$
in Fig.~\ref{fig:split_tree} is not minimal since the purple and gray stars can be joined, creating
the minimal split tree $T_G$.

Cunnigham's definition of a minimal split decomposition is with respect to inclusion. Since
the minimal split decomposition is uniquely determined, it is equivalently the split decomposition
terminating on prime and degenerate graphs using the least number of splits.

\subparagraph*{Computation of Minimal Split Trees.}
The minimal split tree can be computed in time $\calO(n+m)$ using the algorithm of~\cite{dahlhaus}. For
the purpose of this paper, we use the following slower algorithm since it also computes the unique circle
representations of encountered prime circle graphs:

\begin{theorem}[Gioan et al.~\cite{GPTC13b,GPTC13}]
\label{thm:computing_min_decomposition}
The minimal split tree $T_G$ of a circle graph $G$ can be computed in time $\calO((n+m) \cdot
\alpha(n+m))$ where $n$ is the number of vertices, $m$ is the number of edges, and $\alpha$ is the
inverse Ackermann function. Further, the algorithm also computes the unique circle representation of
each prime circle node of $T_G$. 
\end{theorem}

\subparagraph*{Graph Isomorphism Via Minimal Split Decompositions.}
Let $T$ and $T'$ be two graph-labeled trees. An \emph{isomorphism} $\pi : T \to T'$ is an
isomorphism which maps normal edges to normal edges and tree edges to tree edges. Notice that $\pi$
maps nodes of $T$ to isomorphic nodes of $T'$ while preserving tree edges.

We use minimal split trees to test graph isomorphism of circle graphs:

\begin{lemma} \label{lem:minimal_split_tree_isomorphism}
Two connected graphs $G$ and $H$ are isomorphic if and only if the minimal split trees $T_G$ and
$T_H$ are isomorphic.
\end{lemma}

\begin{proof}
Let $T'_G$ and $T'_H$ be any split trees of $G$ and $H$, respectively, and $\pi : T'_G  \to T'_H$ be
an isomorphism. We want to show that $G \cong H$. Choose an arbitrary tree edge $e = m_Am_B$ in
$T'_G$, we know that $\pi(e) = \pi(m_A)\pi(m_B)$ is a tree edge in $T'_H$. We join $T'_G$ over $e$
and $T'_H$ over $\pi(e)$. We get that the restriction $\pi |_{T'_G \setminus \{m_A,m_B\}}$ is an
isomorphism of the constructed graph-labeled trees. By repeating this process, we get single nodes
isomorphic graph-labeled trees which are $G$ and $H$ respectively. So $G \cong H$.

For the other implication, suppose that $\pi : G \to H$ is an isomorphism. Let $D_G$ be a minimal
split decomposition, constructing the minimal split tree $T_G$. We use $\pi$ to construct a split
decomposition $D_H$ and a split tree $T_H$ of $H$ such that $T_G \cong T_H$. Before any splits, the
trees $T_G^0 \cong G$ and $T_H^0 \cong H$ are isomorphic.  Suppose that $T_G^k \cong T_H^k$ and
$D_G$ then uses a split between $A$ and $B$ in some node $N$.  Then $D_H$ will use the split between
$\pi(A)$ and $\pi(B)$, and since $\pi$ is an isomorphism, it is a valid split in $\pi(N)$. We
construct $T_G^{k+1}$ by splitting $N$ into two nodes $N_A$ and $N_B$ and adding marker vertices
$m_A$ and $m_B$, and similarly for $T_H^{k+1}$ with marker vertices $m_{\pi(A)}$ and $m_{\pi(B)}$.
We extend $\pi$ to an isomorphism from $T_G^{k+1}$ to $T_H^{k+1}$ by setting $\pi(m_A) = m_{\pi(A)}$
and $\pi(m_B) = m_{\pi(B)}$. Therefore, the resulting split trees $T_G$ and $T_H$ are isomorphic. By
Theorem~\ref{thm:minimal_decomposition}, the minimal split tree of $H$ is uniquely determined, so
it has to be isomorphic to the constructed $T_H$.
\end{proof}

%% file: canonization_of_trees.tex
\section{Canonization of Graph-labeled Trees} \label{sec:canonization_of_trees}


In the rest of the paper, we work with colored graphs and isomorphisms are required to be
color-preserving. Colors are represented as non-negative integers.

\subparagraph*{Definition of Canonization.}
Let $G$ be a colored graph with $n$ vertices with colors in the range $0,\dots,n-1$ and $m$ edges.
An \emph{encoding} $\eps(G)$ of $G$ is a sequence of non-negative integers. The encoding $\eps(G)$
is \emph{linear} if it contains at most $\calO(n+m)$ integers, each in range $0,\dots,n-1$.  We denote
the class of all encodings by $\calE$.  For a class of graphs $\calC$, a \emph{linear
canonization} is some function $\gamma : \calC \to \calE$ such that $\gamma(G)$ is a linear encoding
of $G$ and for $G,H \in \calC$, we have $G \cong H$ if and only if $\gamma(G) = \gamma(H)$. 

\subparagraph*{Fast Lexicographic Sorting.}
Since we want to sort these encodings lexicographically, we frequently use the following well-known
algorithm:

\begin{lemma}[Aho, Hopcroft, and Ullman~\cite{aho_hopcroft_ullman}, Algorithm 3.2, p. 80]
\label{lem:lex_sort}
It is possible to lexicographically sort sequences of numbers $0,\dots,t-1$ of arbitrary lengths in
time $\calO(\ell + t)$ where $\ell$ is the total length of these sequences.
\end{lemma}

\noindent It is easy to modify the above algorithm to get the same running time when all numbers
belong to $\{0,1,2,\dots,t+1,s,s+1,\dots,s+t-1\}$.

\subparagraph*{Canonization Algorithm.} In the rest of this section, we are going to describe the following
meta-algorithm.

\begin{lemma} \label{lem:glt_can}
Let $\calC$ be a class of graphs and $\calT$ be a class of graph-labeled trees whose nodes belong to
$\calC$. Suppose that we can compute a linear-space canonization $\gamma$ of colored graphs in
$\calC$ in time $f(n+m)$ where $n$ is the number of vertices, $m$ is the number of edges, and $f$ is
convex.  Then we can compute a linear-space canonization $\widetilde\gamma$ of graph-labeled trees
from $\calT$ in time $\calO(n+m+f(n+m))$.
\end{lemma}

Let $T \in \calT$ be a graph-labeled tree. We may assume that $T$ is rooted at the central node. If
a tree edge is central, we insert there another node having a single vertex. Also, we orient all
tree edges towards the root. For a node $N$, we denote by $T[N]$ the graph-labeled subtree induced
by $N$ and all descendants of $N$. Initially, we color all marker vertices by the color $0$ and all
other vertices by the color $1$. Throughout the algorithm, only marker vertices change colors.

The $k$-th \emph{layer} in $T$ is formed by all nodes of the distance $k$ from the root. Notice that
every isomorphism from $T$ to $T'$ maps, for every $k$, the $k$-th layer of $T$ to the $k$-th layer
of $T'$. Also, every node $N$ aside the root is incident to exactly one out-going tree edge whose
incident marker vertex outside $N$ is called the \emph{parent marker vertex of $N$}.

The algorithm starts from the bottom layer of $T$ and process the layers towards the root. When a
node $N$ is processed, we assign a color $c$ to $N$. This color $c$ corresponds to a certain linear
encoding $\gamma'(N)$ which is created by modifying $\gamma(N)$. Further, we store the mapping
$\eps$ from colors to these linear encodings, so $\eps(c) = \gamma'(N)$. The assigned colors have
the property that two nodes $N$ and $N'$ have the same assigned color if and only if the rooted
graph-labeled subtrees induced by $N$ and all the nodes below and by $N'$ and all the nodes below,
respectively, are isomorphic. We remove the node $N$ from $T$ and assign its color to the parent
marker vertex of $N$. Also notice that when a non-root node is processed, all marker vertices except
for one have colors different from $0$, and for the root node, all marker vertices have colors
different from $0$.

Let $N_1,\dots,N_k$ be the nodes of the currently processed layer such that each vertex in these
nodes has one color from $\{0,1,s,s+1,\dots,s+t-1\}$. For each node $N_i$, we want to use the
canonization subroutine to compute the linear encoding $\gamma(N_i)$. But the assumptions require
that for $\ell$ vertices in $N_i$, all colors are in range $0,\dots,\ell-1$, but we might have $s
\gg \ell$. We can avoid this by renumbering the colors since at most $\ell$ different colors are
used on $N_i$. Suppose that exactly $c_i$ different colors are used in $N_i$, and we define the
injective mapping $\varphi_{N_i} : \{0,1,\dots,c_i-1\} \to \{0,1,s,s+1,\dots,s+t-1\}$ such that the
smallest used color is $\varphi_{N_i}(0)$, the second smallest is $\varphi_{N_i}(1)$, and so on till
$\varphi_{N_i}(c_i-1)$. The renumbering of colors on $N_i$ is given by the inverse
$\varphi_{N_i}^{-1}$.

After renumbering, the algorithm runs the canonization subroutine to compute the linear encodings
$\gamma(N_1),\dots,\gamma(N_k)$. We create the modified linear encoding $\gamma'(N_i)$ by
pre-pending $\gamma(N_i)$ with the sequence $c_i$, $\varphi_{N_i}(0)$, $\varphi_{N_i}(1)$, \dots,
$\varphi_{N_i}(c_i-1)$ where $c_i$ is the number of different colors used in $N_i$.  The algorithm
lexicographically sorts these modified encodings $\gamma'(N_1),\dots,\gamma'(N_k)$ using
Lemma~\ref{lem:lex_sort}. Next, we assign the color $s+t$ to the nodes having the smallest
encodings, the color $s+t+1$ to the nodes having the second smallest encodings, and so on. For
every node $N_i$, we remove it and set the color of the parent marker vertex of $N_i$ to the color
assigned to $N_i$.

Suppose that the root node $N$ has the color $c$ assigned, so throughout the algorithm, we have used
the colors $0,\dots,c$. The computed linear encoding of $T$ is an encoded concatenation of
$\eps(2),\eps(3),\dots,\eps(c)$.

\begin{lemma} \label{lem:glt_can_correctness}
The described algorithm produces a correct linear canonization $\widetilde\gamma$ of graph-labeled
trees in $\calT$, i.e., for $T,T' \in \calT$, we have $T \cong T'$ if and only if
$\widetilde\gamma(T) = \widetilde\gamma(T')$.
\end{lemma}

\begin{proof}
Let $\pi : T \to T'$ be an isomorphism. We prove by induction from the bottom layer to the root that
$\gamma'(N) = \gamma'(\pi(N))$. Suppose that we are processing the $\ell$-th layer of $T$ and $T'$
and all previously used colors have the same assigned encodings in $T$ and $T'$. For every node $N$
of the $\ell$-th layer of $T$, then $\pi(N)$ belongs to the $\ell$-th layer of $T'$. We argue that
$\pi$ also preserves the colors of $N$. Since $\pi$ is an isomorphism of graph-labeled trees, it
maps marker vertices to marker vertices and non-marker vertices to non-marker vertices. Let
$N_1,\dots,N_k$ be the children of $N$. By the induction hypothesis, we have $\gamma'(N_i) =
\gamma'(\pi(N_i))$, so the same colors are assigned to $N_i$ and $\pi(N_i)$. Therefore, $\pi$
preserves the colors of $N$, and this property still holds after renumbering the colors by
$\varphi_N^{-1} = \varphi_{\pi(N)}^{-1}$. Therefore, we have $\gamma(N)=\gamma(\pi(N))$ and thus
$\gamma'(N)=\gamma'(\pi(N))$. Thus, the lexicographic sorting of the nodes in the $\ell$-th layer of
$T$ and of $T'$ is the same, so $N$ and $\pi(N)$ have the same color assigned. Finally, the
encodings $\eps(2),\dots,\eps(c)$ are the same in $T$ and $T'$, so $\widetilde\gamma(T) =
\widetilde\gamma(T')$.

For the other implication, we show that a graph-labeled tree $T'$ isomorphic to $T$ can be
reconstructed from $\widetilde\gamma(T)$. We construct the root node $N$ from $\eps(c)$. Since
$\gamma$ is a canonization of $\calC$, we obtain $N$ by applying $\gamma^{-1}$. Next, we invert the
recoloring by applying $\varphi(N)$ on the colors of $N$. Next, we consider each marker vertex in
$N$. If it has some color $c_i$, we use $\eps(c_i)$ to construct a child node $N'$ of $N$ exactly as
before. We proceed in this way till all nodes are expanded and only the colors $0$ and $1$ remain.
It is easy to prove by induction that $T \cong T'$ since each $\eps(c_i)$ uniquely determines the
corresponding subtree in $T$.
\end{proof}

\begin{lemma} \label{lem:glt_can_complexity}
The described algorithm runs in time $\calO(n+m+f(n+m))$.
\end{lemma}

\begin{proof}
When we run the canonization subroutine on a node having $n'$ vertices and $m'$ edges, it has all colors in
range $0,\dots,n'-1$, so we can compute its linear encoding in time $f(n'+m')$. Since $f$ is convex
and the canonization subroutine runs on each node exactly once, the total time spend by this
subroutine is bounded by $f(n+m)$.

The total count of used colors is clearly bounded by $n$ and each layer uses different colors except
for $0$ and $1$. Consider a layer with nodes $N_1,\dots,N_k$ having $\ell$ vertices and $\ell'$
edges in total. All these nodes use at most $\ell$ different colors. Therefore, the modified
encodings $\gamma'(N_1),\dots,\gamma'(N_k)$ consisting of integers from
$\{0,1,2,\dots,\ell+1,s,s+1,\dots,s+\ell-1\}$, for some value $s$, and are of the total length
$\calO(\ell+\ell')$.  Therefore, lexicografic sorting of these modified encodings can be done in time
$\calO(\ell+\ell')$, and this sorting takes total time $\calO(n+m)$ for all layers of $T$. Therefore,
the total running time of the algorithm is $\calO(n+m+f(n+m))$.
\end{proof}

\begin{proof}[Proof of Lemma~\ref{lem:glt_can}]
It follows by Lemmas~\ref{lem:glt_can_correctness} and~\ref{lem:glt_can_complexity}.
\end{proof}

%% file: canonization_of_nodes.tex
\section{Canonization of Prime and Degenerate Circle Graphs} \label{sec:canonization_of_nodes}

Let $\calC$ be the class of colored prime and degenerate circle graphs. Recall that all nodes of
minimal split trees of connected circle graphs belong to $\calC$. To apply the meta-algorithm of
Lemma~\ref{lem:glt_can}, we need to show that the linear canonization $\gamma$ of $\calC$ can be
computed in time $\calO(n+m)$ where $n$ is the number of vertices and $m$ is the number of edges.

\subparagraph*{Linear Canonizations of Degenerate Graphs.}
For a colored complete graph $G = K_n$, we sort its colors using bucket sort in time $\calO(n)$, so
the vertices have the colors $c_1 \le c_2 \le \dots \le c_n$. The computed linear canonization
$\gamma(G)$ is $0,c_1,c_2,\dots,c_n$. 

For a star $G = S_n$, we sort the colors of leaves using bucket sort in time $\calO(n)$, so they
have the colors $c_1 \le c_2 \le \dots \le c_n$, while the center has the color $c_0$. The computed
linear canonization $\gamma(G)$ is $1,c_0,c_1,c_2,\dots,c_n$.

\subparagraph*{Linear Encodings of Colored Cycles.}
As a subroutine, we need to find a canonical form  of a colored cycle.
To do this, it suffices to find the the lexicographically minimal rotation of a circular string.
This can be done using $\calO(n)$ comparisons over some alphabet $\Sigma$~\cite{booth1980lexicographically,shiloach1981fast}.

\subparagraph*{Linear Canonizations of Circle Representations.}
Let $G$ be an arbitrary colored circle graph on $n$ vertices together with an arbitrary circle
representation $\calR$. The standard way to describe $\calR$ is to arbitrarily order the vertices
$1,\dots,n$ and to give a circular word $\omega$ consisting of $2n$ integers from $1,\dots,n$, each
appearing exactly twice, in such a way that the occurrences of $i$ and $j$ alternate (i.e., appear as
$ijij$) if and only if $ij \in E(G)$. This circular word describes the ordering of endpoint of the
chords in, say, the clockwise direction.

Let $G$ and $H$ be two colored circle graphs on $n$ vertices labeled $1,\dots,n$ with circle
representations $\calR_G$ and $\calR_H$ represented by $\omega_G = \omega_G^1,\dots,\omega_G^{2n}$
and $\omega_H = \omega_H^1,\dots,\omega_H^{2n}$. We say that $\calR_G \cong \calR_H$ if and only if
there exists a bijection $\pi : \{1,\dots,n\} \to \{1,\dots,n\}$ such that
\begin{itemize}
\item the vertices $i$ in $G$ and $\pi(i)$ in $H$ have identical colors, and
\item the circular words $\omega_H$ and $\pi(\omega_G^1),\dots,\pi(\omega_G^{2n})$ are identical.
\end{itemize}
Notice that when $\calR_G \cong \calR_H$, necessarily $G \cong H$, but in general the converse is
not true.  We want to construct a linear canonization $\gamma$ such that $\calR_G \cong \calR_H$ if
and only if $\gamma(\calR_G) = \gamma(\calR_H)$.

We want to construct $\gamma(\calR_G)$ directly from the representation but an arbitrary chosen
labeling of vertices in $\omega_G$ is not helpful. Therefore, we consider a different encoding of
the representation which is invariant on rotation. For each of $2n$ endpoints $e_1,\dots,e_{2n}$, we
store two numbers:
\begin{itemize}
\item The color $c_i \in \{0,\dots,n-1\}$ of the vertex of the chord corresponding to $e_i$.
\item The number of endpoints $g_i$ in the clockwise direction between $e_i$ and the other endpoint
corresponding to the same chord. We have $g_i \in \{0,\cdots,2n-2\}$ and when the $e_i$ and $e_j$
correspond one chord, then $g_i+g_j = 2n-2$.
\end{itemize}
To distinguish $c_i$ from $g_i$, we increase all $c_i$ by $2n-1$, so $c_i \in \{2n-1,\cdots,3n-2\}$.
Then we may consider the circular word $\lambda_G = g_1,c_1,g_2,c_2,\dots,g_{2n},c_{2n}$ of length
$4n$.

\begin{lemma} \label{lem:circle_rep_encoding}
Let $G$ and $H$ be two colored circle graphs with representations $\calR_G$ and $\calR_H$. We have
$\calR_G \cong \calR_H$ if and only if the circular words $\lambda_G$ and $\lambda_H$ are
identical.
\end{lemma}

\begin{proof}
If $\calR_G \cong \calR_H$, there exists an index $k \in \{0,\dots,2n-1\}$ such that rotating the
representation $\calR_G$ by $k$ endpoints produces $\calR_H$. When $\lambda_G =
g_1,c_1,\dots,g_{2n},c_{2n}$ and $\lambda_H = g'_1,c'_1,\dots,g'_{2n},c'_{2n}$, it cyclically
holds that $g_i = g'_{i+k}$ and $c_i = c'_{i+k}$. So the circular words $\lambda_G$ and
$\lambda_H$ are identical.

For the other implication, observe that the circle representation $\calR_G$ and the circle graph $G$
can be reconstructed from $\lambda_G$. If $\lambda_G$ and $\lambda_H$, we reconstruct isomorphic
representations $\calR_G$ and $\calR_H$.
\end{proof}

\begin{lemma} \label{lem:circle_rep_can}
We can compute the linear encoding $\gamma$ of colored circle representations in time $\calO(n)$.
\end{lemma}

\begin{proof}
For a representation $\calR_G$, we can clearly compute $\lambda_G$ in time $\calO(n)$. Next, we apply
to $\lambda_G$ an cycle canonization algorithm~\cite{booth1980lexicographically,shiloach1981fast}  which computes
$\gamma(\calR_G)$ in time $\calO(n)$.
\end{proof}

\subparagraph*{Linear Canonization of Prime Circle Graphs.} Let $G$ be a prime circle graph. It has
at most two different representations $\calR_G$ and $\calR'_G$ where one is the reversal of the
other. Using Lemma~\ref{lem:circle_rep_can}, we compute their linear encodings $\lambda_G$ and
$\lambda'_G$. As the linear encoding $\gamma(G)$, we chose the lexicographically smallest of
$\lambda_G$ and $\lambda'_G$, prepended with the value $2$. Clearly, colored prime circle graphs $G$
and $H$ are isomorphic if and only if $\gamma(G) = \gamma(H)$.

By putting the results of this section together, we get the following:

\begin{lemma} \label{lem:circle_node_can}
We can compute linear canonization of colored prime circle graphs and degenerate graphs in time
$\calO(n)$.
\end{lemma}

%% file: canonization_of_circle_graphs.tex
    \section{Canonization and Graph Isomorphism of Circle Graphs}
\label{sec:canonization_of_circle_graphs}

In this section, we combine the presented results to show that a linear canonization $\gamma$ of
circle graphs can be computed in time $\calO((n+m) \cdot \alpha(n+m))$. This algorithm clearly implies
Theorem~\ref{thm:graphiso} since circle graphs $G$ and $H$ are isomorphic if and only if
$\gamma(G)=\gamma(H)$.

Suppose that $G$ is a connected circle graph. We apply the algorithm of
Theorem~\ref{thm:computing_min_decomposition} to compute the minimal split decomposition $T_G$ of
$G$ and the unique circle representation for each prime circle graph (up to reversal). We halt if
some circle representations does not exist since $G$ is not a circle graph.  The total running time
of preprocessing is $\calO((n+m) \cdot \alpha(n+m))$, and the remainder of the algorithm runs in time
$\calO(n+m)$, so this step is the bottleneck. Next, we use Lemmas~\ref{lem:circle_node_can}
and~\ref{lem:glt_can} to compute a linear canonization $\gamma(T_G)$ and we put $\gamma(G) =
\gamma(T_G)$.

Suppose that the circle graph $G$ is disconnected, and let $G_1,\dots,G_k$ be its connected
components. We compute their linear encodings $\gamma(G_1),\dots,\gamma(G_k)$, lexicografically sort
them in time $\calO(n+m)$ using Lemma~\ref{lem:lex_sort}, and output them in $\gamma(G)$ sorted as a
sequence. The total running time is $\calO((n+m) \cdot \alpha(n+m))$.

When the input also gives a circle representation $\calR$, we can avoid using
Theorem~\ref{thm:computing_min_decomposition}. Instead, we compute a split decomposition and the
corresponding split tree in time $\calO(n+m)$ using~\cite{dahlhaus}. We can easily modify this split
tree into the minimal split tree by joining neighboring complete vertices and stars as discussed in
Section~\ref{sec:split_decomposition}. For each prime node $N$, we obtain its unique circle
representation by restricting $\calR$ to the vertices of $N$. Since the avoided algorithm of
Theorem~\ref{thm:computing_min_decomposition} was the bottleneck, we get the total running time
$\calO(n+m)$.

%% file: conclusions.tex
\section{Conclusions} \label{sec:conclusions}

We conclude this paper by discussing several possible research directions and open problems. The
main used tool is minimal split decomposition. The algorithm finds the cannonical form of the unique
minimal split tree $T$, using the canonical forms of every prime and degenerate circle graph
appearing as a node of $T$. We obtain an algorithm computing the cannonical form for every circle
graph.

\begin{problem}
Does the minimal split tree capture all possible representations of circle graph?
\end{problem}

The \emph{$k$-dimensional Weisfeiler-Leman}~\cite{wl} algorithm ($k$-dim WL) is a fundamental
algorithm used as a subroutine in graph isomorphism testing.  The algorithm colors $k$-tuples of
vertices of two input graphs and iteratively refines the color classes until the coloring becomes stable.
We say that $k$-dim WL \emph{distinguishes} two graph $G$ and $H$ if and only if its application to each of
them gives colorings with different sizes of color classes. Two distinguished graphs are clearly non-isomorphic,
however, for every $k$, there exist non-isomorphic graphs not distinguished by $k$-dim WL.
For a class of graphs $\calC$, the \emph{Weisfeiler-Leman dimension} of $\calC$ is the
minimum integer $k$ such that $k$-dim WL distinguishes every $G,H \in \calC$ such that $G \not\cong H$.

\begin{problem}
What is the Weisfeiler-Leman dimension of circle graphs?
\end{problem}

For many classes of graphs, such as trees, interval graphs~\cite{evdokimov2000forestal}, planar graphs~\cite{kiefer2017weisfeiler}, or more
generally graphs with excluded minors~\cite{grohe2017descriptive}, and recently also circular-arc graphs~\cite{nedela2019testing}, the
Weisfeiler-Leman dimension is finite.  A major open problem is for which classes of graphs can
isomorphism be tested in polynomial time without using group theory, i.e., by a combinatorial
algorithm, often meant some $k$-dim WL.

Further, a natural question to ask, especially for problems solvable in linear time, is whether
they can be solved using logarithmic space. This is, for example, known for interval graphs~\cite{kobler2011interval} and Helly circular-arc graphs~\cite{kobler2013helly}.
Recently, a parametrized logspace algorithm was given for circular-arc graphs in~\cite{chandoo2016deciding}.

\begin{problem}
Can isomorphism of circle graphs be tested using logarithmic space?
\end{problem}

Finally, we mention the \emph{partial representation extension} problem which is a generalization of
the recognition problem. The input consists of, in our case, a circle graph and a circle
representation of its induced subgraph and the task is to complete the representation or output that
it is not possible.  Obviously, this problem can be asked for various classes of graphs and it was
extensively studied~\cite{kkosv,kkorssv17,kkos15,cfk15,jensen,trapezoid_repext}.

In~\cite{cfk15}, the authors give an $\calO(n^3)$ algorithm to solve this problem. They give an
elementary recursive description of the structure of all representations and try to match the
partial representation on it.  It is a natural question whether the minimal split trees can be used
to solve the partial representation extension problem faster.

\begin{problem}
Can the partial representation extension problem for circle graphs be solved using minimal split
decomposition faster that $\calO(n^3)$?
\end{problem}

%% file: circle_graph_iso.bbl
\begin{thebibliography}{10}

\bibitem{aho_hopcroft_ullman}
A.~V. Aho, J.~E. Hopcroft, and J.~D. Ullman.
\newblock {\em The Design and Analysis of Computer Algorithms}.
\newblock Addison-Wesley Publishing Company, 1974.

\bibitem{babai_quasipoly}
L.~Babai.
\newblock Graph isomorphism in quasipolynomial time.
\newblock In {\em STOC}, 2016.

\bibitem{jensen}
J.~Bang{-}Jensen, J.~Huang, and X.~Zhu.
\newblock Completing orientations of partially oriented graphs.
\newblock {\em CoRR}, abs/1509.01301, 2015.

\bibitem{booth1980lexicographically}
Kellogg~S Booth.
\newblock Lexicographically least circular substrings.
\newblock {\em Information Processing Letters}, 10(4-5):240--242, 1980.

\bibitem{B1987}
A.~Bouchet.
\newblock Reducing prime graphs and recognizing circle graphs.
\newblock {\em Combinatorica}, 7(3):243--254, 1987.

\bibitem{B87}
A.~Bouchet.
\newblock Unimodularity and circle graphs.
\newblock {\em Discrete Mathematics}, 66(1-2):203--208, 1987.

\bibitem{chandoo2016deciding}
Maurice Chandoo.
\newblock Deciding circular-arc graph isomorphism in parameterized logspace.
\newblock In {\em 33rd Symposium on Theoretical Aspects of Computer Science
  (STACS 2016)}. Schloss Dagstuhl-Leibniz-Zentrum fuer Informatik, 2016.

\bibitem{cfk15}
S.~Chaplick, R.~Fulek, and P.~Klav\'{\i}k.
\newblock Extending partial representations of circle graphs.
\newblock {\em CoRR}, abs/1309.2399, 2015.

\bibitem{permutation_isomorphism}
C.~J. Colbourn.
\newblock On testing isomorphism of permutation graphs.
\newblock {\em Networks}, 11(1):13--21, 1981.

\bibitem{cunningham1982decomposition}
William~H Cunningham.
\newblock Decomposition of directed graphs.
\newblock {\em SIAM Journal on Algebraic Discrete Methods}, 3(2):214--228,
  1982.

\bibitem{cunningham1980combinatorial}
William~H Cunningham and Jack Edmonds.
\newblock A combinatorial decomposition theory.
\newblock {\em Canadian Journal of Mathematics}, 32(3):734--765, 1980.

\bibitem{isomorphism_circular_one_property}
A.~R. Curtis, M.~C. Lin, R.~M. McConnell, Y.~Nussbaum, F.~J. Soulignac, J.~P.
  Spinrad, and J.~L. Szwarcfiter.
\newblock Isomorphism of graph classes related to the circular-ones property.
\newblock {\em Discrete Mathematics and Theoretical Computer Science},
  15(1):157--182, 2013.

\bibitem{dahlhaus}
E.~Dahlhaus.
\newblock Parallel algorithms for hierarchical clustering and applications to
  split decomposition and parity graph recognition.
\newblock {\em Journal of Algorithms}, 36(2):205--240, 1998.

\bibitem{deF81}
H.~de~Fraysseix.
\newblock Local complementation and interlacement graphs.
\newblock {\em Discrete Mathematics}, 33(1):29--35, 1981.

\bibitem{deFM99}
H.~de~Fraysseix and P.~O. de~Mendez.
\newblock On a characterization of gauss codes.
\newblock {\em Discrete {\&} Computational Geometry}, 22(2):287--295, 1999.

\bibitem{evdokimov2000forestal}
Sergei Evdokimov, Ilia Ponomarenko, and Gottfried Tinhofer.
\newblock Forestal algebras and algebraic forests (on a new class of weakly
  compact graphs).
\newblock {\em Discrete Mathematics}, 225(1-3):149--172, 2000.

\bibitem{EI71}
S.~Even and A.~Itai.
\newblock Queues, stacks, and graphs.
\newblock {\em {T}heory of {M}achines and {C}omputation (Z. Kohavi and A. Paz,
  Eds.)}, pages 71--76, 1971.

\bibitem{fkkn16}
J.~Fiala, P.~Klav\'{\i}k, J.~Kratochv\'{\i}l, and R.~Nedela.
\newblock 3-connected reduction for regular graph covers.
\newblock {\em CoRR}, abs/1503.06556, 2017.

\bibitem{GSH89}
C.~P. Gabor, K.~J. Supowit, and W.~Hsu.
\newblock Recognizing circle graphs in polynomial time.
\newblock {\em J. ACM}, 36(3):435--473, 1989.

\bibitem{GPTC13}
E.~Gioan, C.~Paul, M.~Tedder, and D.~Corneil.
\newblock Practical and efficient circle graph recognition.
\newblock {\em Algorithmica}, 69(4):759--788, 2014.

\bibitem{GPTC13b}
E.~Gioan, C.~Paul, M.~Tedder, and D.~Corneil.
\newblock Practical and efficient split decomposition via graph-labelled trees.
\newblock {\em Algorithmica}, 69(4):789--843, 2014.

\bibitem{grohe2017descriptive}
Martin Grohe.
\newblock {\em Descriptive complexity, canonisation, and definable graph
  structure theory}, volume~47.
\newblock Cambridge University Press, 2017.

\bibitem{hopcroft1974linear}
J.~E. Hopcroft and J.~Wong.
\newblock Linear time algorithm for isomorphism of planar graphs.
\newblock In {\em STOC}, pages 172--184. ACM, 1974.

\bibitem{hsu1995m}
W.~L. Hsu.
\newblock {$O(M \cdot N)$} algorithms for the recognition and isomorphism
  problems on circular-arc graphs.
\newblock {\em SIAM Journal on Computing}, 24(3):411--439, 1995.

\bibitem{kkmnz}
K.~Kawarabayashi, P.~Klav\'{\i}k, B.~Mohar, R.~Nedela, and P.~Zeman.
\newblock Isomorphisms of maps on the sphere.
\newblock {\em to appear in Contemporary Mathematics AMS}, 2019.

\bibitem{kiefer2017weisfeiler}
Sandra Kiefer, Ilia Ponomarenko, and Pascal Schweitzer.
\newblock The weisfeiler-leman dimension of planar graphs is at most 3.
\newblock In {\em 2017 32nd Annual ACM/IEEE Symposium on Logic in Computer
  Science (LICS)}, pages 1--12. IEEE, 2017.

\bibitem{kkz}
P.~Klav\'{\i}k, D.~Knop, and P.~Zeman.
\newblock Graph isomorphism restricted by lists.
\newblock {\em CoRR}, abs/1607.03918, 2016.

\bibitem{kkorssv17}
P.~Klav\'{\i}k, J.~Kratochv\'{\i}l, Y.~Otachi, I.~Rutter, T.~Saitoh,
  M.~Saumell, and T.~Vysko\v{c}il.
\newblock Extending partial representations of proper and unit interval graphs.
\newblock {\em Algorithmica}, 77(4):1071--1104, 2017.

\bibitem{kkos15}
P.~Klav\'{\i}k, J.~Kratochv\'{\i}l, Y.~Otachi, and T.~Saitoh.
\newblock Extending partial representations of subclasses of chordal graphs.
\newblock {\em Theoretical Computer Science}, 576:85--101, 2015.

\bibitem{kkosv}
P.~Klav\'{\i}k, J.~Kratochv\'{\i}l, Y.~Otachi, T.~Saitoh, and T.~Vysko\v{c}il.
\newblock Extending partial representations of interval graphs.
\newblock {\em Algorithmica}, 2016.

\bibitem{kz}
P.~Klav\'{\i}k and P.~Zeman.
\newblock Automorphism groups of geometrically represented graphs.
\newblock In {\em 32nd International Symposium on Theoretical Aspects of
  Computer Science, STACS 2015}, volume~30 of {\em Leibniz International
  Proceedings in Informatics (LIPIcs)}, pages 540--553, 2015.

\bibitem{kobler2011interval}
Johannes K{\"o}bler, Sebastian Kuhnert, Bastian Laubner, and Oleg Verbitsky.
\newblock Interval graphs: Canonical representations in logspace.
\newblock {\em SIAM Journal on Computing}, 40(5):1292--1315, 2011.

\bibitem{kobler2013helly}
Johannes K{\"o}bler, Sebastian Kuhnert, and Oleg Verbitsky.
\newblock Helly circular-arc graph isomorphism is in logspace.
\newblock In {\em International Symposium on Mathematical Foundations of
  Computer Science}, pages 631--642. Springer, 2013.

\bibitem{trapezoid_repext}
T.~Krawczyk and B.~Walczak.
\newblock Extending partial representations of trapezoid graphs.
\newblock In {\em WG 2017}, Lecture Notes in Computer Science, 2017.

\bibitem{lin2008simple}
Min~Chih Lin, Francisco~J Soulignac, and Jayme~L Szwarcfiter.
\newblock A simple linear time algorithm for the isomorphism problem on proper
  circular-arc graphs, 2008.

\bibitem{lueker1979linear}
G.~S. Lueker and K.~S. Booth.
\newblock A linear time algorithm for deciding interval graph isomorphism.
\newblock {\em Journal of the ACM (JACM)}, 26(2):183--195, 1979.

\bibitem{N85}
W.~Naji.
\newblock {\em Graphes de Cordes: Une Caracterisation et ses Applications}.
\newblock PhD thesis, l'Universit\'{e} Scientifique et M\'{e}dicale de
  Grenoble, 1985.

\bibitem{nedela2019testing}
Roman Nedela, Ilia Ponomarenko, and Peter Zeman.
\newblock Testing isomorphism of circular-arc graphs in polynomial time.
\newblock {\em arXiv preprint arXiv:1903.11062}, 2019.

\bibitem{oum}
S.~Oum.
\newblock Rank-width and vertex-minors.
\newblock {\em J. Comb. Theory, Ser. B}, 95(1):79--100, 2005.

\bibitem{schoning1988graph}
U.~Sch{\"o}ning.
\newblock Graph isomorphism is in the low hierarchy.
\newblock {\em Journal of Computer and System Sciences}, 37(3):312--323, 1988.

\bibitem{shiloach1981fast}
Yossi Shiloach.
\newblock Fast canonization of circular strings.
\newblock {\em Journal of algorithms}, 2(2):107--121, 1981.

\bibitem{S94}
J.~P. Spinrad.
\newblock Recognition of circle graphs.
\newblock {\em J. of Algorithms}, 16(2):264--282, 1994.

\bibitem{egr}
J.~P. Spinrad.
\newblock {\em Efficient Graph Representations}.
\newblock Field Institute Monographs, 2003.

\bibitem{wl}
B.~Weisfeiler and A.A. Leman.
\newblock A reduction of a graph to a canonical form and an algebra arising
  during this reduction.
\newblock {\em Nauchno-Technicheskaya Informatsiya}, 9:12--16, 1968.

\end{thebibliography}
